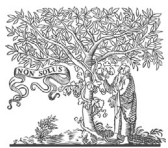

Contents lists available at ScienceDirect

# MethodsX

journal homepage: www.elsevier.com/locate/mex

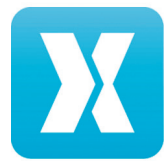

Method Article

# Wavelet regression: An approach for undertaking multi-time scale analyses of hydro-climate relationships

## Jianhua Xu[a,b,*]

[a] *Key Laboratory of Geographic Information Science (Ministry of Education), School of Geographic Sciences, East China Normal University, Shanghai, 200241, China*
[b] *Research Center for East-West Cooperation in China, East China Normal University, Shanghai, 200241, China*

A B S T R A C T

Previous studies showed that hydro-climate processes are stochastic and complex systems, and it is difficult to discover the hidden patterns in the all non-stationary data and thoroughly understand the hydro-climate relationships. For the purpose to show multi-time scale responses of a hydrological variable to climate change, we developed an integrated approach by combining wavelet analysis and regression method, which is called wavelet regression (WR). The customization and the advantage of this approach over the existing methods are presented below:

- The patterns in the data series of a hydrological variable and its related climatic factors are revealed by the wavelet analysis at different time scales.
- The hydro-climate relationship of each pattern is revealed by the regression method based on the results of wavelet analysis.
- The advantage of this approach over the existing methods is that the approach provides a routing to discover the hidden patterns in the stochastic and non-stationary data and quantitatively describe the hydro-climate relationships at different time scales.

* Corresponding author at: Research Center for East-West Cooperation in China, East China Normal University, Shanghai, 200241, China.
   *E-mail address:* jhxu@geo.ecnu.edu.cn (J. Xu).

## Wavelet analysis

Wavelet transformation has been shown to be a powerful technique for characterizing the frequency, intensity, scale, and duration of variations in hydro-climatic process [1–3]. Wavelet analysis can also reveal localized time and frequency information without requiring the signal time series to be stationary, as required by the Fourier transform and other spectral methods [4].

The first task of WR is to approximate the variation patterns of a hydrological variable and its related climatic factors by using wavelet decomposition and reconstruction at different time scales.

The principle of wavelet decomposition and reconstruction is as follows [5,6]. Considering a given signal $X(t)$, such as streamflow, temperature and precipitation, etc., which can be built up as a sequence of projections onto Father and Mother wavelets indexed by both $k$ {$k = 1, 2, \ldots \ldots$ } and $s$ {$s = 2^j$, $j = 1, 2, \ldots \ldots$ }. The coefficients in the expansion are given by the projections

$$\begin{cases} s_{J,k} = \int X(t)\Phi_{J,k}(t)dt \\ d_{j,k} = \int X(t)\Psi_{j,k}(t) j = 1,2,\cdots,J \end{cases} \tag{1}$$

where $J$ is the maximum scale sustainable by the number of data points, $\Phi_{j,k}(t) = 2^{-\frac{j}{2}}\Phi(\frac{t-2^j k}{2^j})$ is father wavelet, and $\Psi_{j,k}(t) = 2^{-\frac{j}{2}}\Psi(\frac{t-2^j k}{2^j})$ is mother wavelet. Generally, father wavelet is used for the lowest-frequency smooth components, which requires wavelet with the widest support; mother wavelet is used for the highest-frequency detailed components. In other words, father wavelet is used for the major trend components, and mother wavelet is used for all deviations from the trend [6,7].

Once a mother wavelet is selected, the wavelet transform can be used to decompose a signal according to scale, allowing separation of the fine-scale behavior (detail) from the large-scale behavior (approximation) of the signal [6,7]. The relationship between scale and signal behavior is designated as follows: a low scale corresponds to compressed wavelet as well as rapidly changing details, namely high frequency, whereas a high scale corresponds to stretched wavelet and slowly changing coarse features, namely low frequency. Signal decomposition is typically conducted in an iterative fashion using a series of scales such as $a = 2, 4, 8, \ldots \ldots, 2^L$, with successive approximations being split in turn so that one signal is broken down into many lower resolution components [6].

The representation of the signal $X(t)$ now can be given by:

$$X(t) = S_J + D_J + D_{J-1} + \ldots + D_j + \ldots + D_1 \tag{2}$$

where $S_J = \sum_k s_{J,k}\Phi_{J,k}(t)$ and $D_j = \sum_k d_{j,k}\Psi_{j,k}(t)$, $j = 1, 2, \ldots, J$.

In general, we have the relationship as

$$S_{j-1} = S_j + D_j \tag{3}$$

where {$S_J, S_{J-1}, \ldots, S_1$}is a sequence of multi-resolution approximations of the function $X(t)$ at ever-increasing levels of refinement. The corresponding multi-resolution decomposition of $X(t)$ is given by {$S_J$, $D_J$, $D_{J-1}$, $\ldots \ldots$, $D_j$, $\ldots \ldots$, $D_1$} [6].

Selecting a proper wavelet function is a prerequisite for wavelet analysis. The actual criteria for wavelet selection include self-similarity, compactness, and smoothness [8,9]. Choosing the Symmlet family [6], we experimented with alternative choices of scaling functions, and found that the results from 'Sym8' are robust. Therefore, 'sym8' is used for approximating the variation patterns of the hydrological variable and its related climatic factors at different time scales.

## Regression analysis based on the results of wavelet analysis

The second task of WR is to fit the regression equation describing the hydro-climate relationship between a hydrological variable with its related climatic factors for each pattern at the chosen time scale based on the results of wavelet analysis.

Because the hydrological variable (e.g. streamflow or groundwater table, etc.) is affected by climatic factors (e.g. temperature, precipitation, etc.), we generally suppose the hydrological variable as dependent variable, $Y$, and climatic factors as independent variables, $X_1$, $X_2$, $\ldots$, $X_k$. The linear regression equation is as follows:

$$Y = b_0 + b_1X_1 + b_2X_2 + \ldots + b_kX_k \tag{4}$$

where, $b_0$ is a constant, and $b_1$, $b_2$, ...., and $b_k$ are partial regression coefficient, which can be fitted by the Least Squares [10]. The significance of the regression Eq. (4) should be tested by the F-test with a significance level [11].

If the linear regression equation cannot well describe the hydro-climate relationship between a hydrological variable with its related climatic factors, we must fit a nonlinear regression equation instead.

Using the above approach, we can fit a most suitable regression equation for each pattern at the chosen time scale to describe the hydro-climate relationship between a hydrological variable with its related climatic factors based on the results of wavelet analysis. We called the above regression equation based on the results of wavelet analysis as wavelet regression equation (WRE).

**The test of wavelet regression equation**

We suggest to use the coefficient of determination and Akaike information criterion (AIC) to test the fitting effect of the above WRE.

In order to identify the uncertainty of the wavelet regression equation for a given time scale, the coefficient of determination, also known as the goodness of fit, was calculated as follows [11]:

$$R^2 = 1 - \frac{RSS}{TSS} = 1 - \frac{\sum_i^n (Y_i - \hat{Y}_i)^2}{\sum_i^n (Y_i - \overline{Y})^2} \tag{5}$$

where $R^2$ is the coefficient of determination; $\hat{Y}_i$ and $Y_i$ are the simulate value by the WRE and observed data of the hydrological variable; $\overline{Y}$ is the mean of $Y_i$ (i = 1, 2, ......, n); $RSS = \sum_{i=1}^n (Y_i - \hat{Y}_i)^2$ is the residual sum of squares; $TSS = \sum_{i=1}^n (Y_i - \overline{Y})^2$ is the total sum of squares.

The coefficient of determination is a measure of how well the simulate results represent the actual data. A bigger $R^2$ indicates a higher certainty and lower uncertainty of the WRE.

To compare the relative goodness between different WREs, we also used the measure of Akaike information criterion (AIC) [12]. The formula of AIC is as follows:

$$AIC = 2k + \ln(RSS/n) \tag{6}$$

where k is the number of parameters estimated in the model; n is the number of samples; $RSS$ is the same as in formula (5). A smaller AIC indicates a better goodness of the WRE [12].

For small sample sizes (i.e., $n/k \leq 40$), the second-order Akaike Information Criterion ($AIC_c$) should be used instead

$$AIC_c = AIC + \frac{2k(k+1)}{n-k-1} \tag{7}$$

where n is the sample size. As the sample size increases, the last term of the $AIC_c$ approaches zero, and the $AIC_c$ tends to yield the same conclusions as the AIC [13].

**An application case**

As known, the observation data from hydrological and climatic stations always present the stochastic and non-stationary characteristic. How can we reveal the patterns hidden in the stochastic and non-stationary data? Several application cases [14–17] in Northwest China have proved that the WR is effective approach, which present a good performance.

Fig.1 shows the observation data of annual average temperature (AAT), annual precipitation (AP) and annual runoff (AR) in the Yarkand River basin of Northwest China. It is evident that all the data series of AAT, AP and AR are fluctuating, and difficult to identify the patterns hidden in the raw data. In order to uncover the patterns hidden in the raw data, we now analyze hydro-climate relationships at different time scales using the wavelet regression.

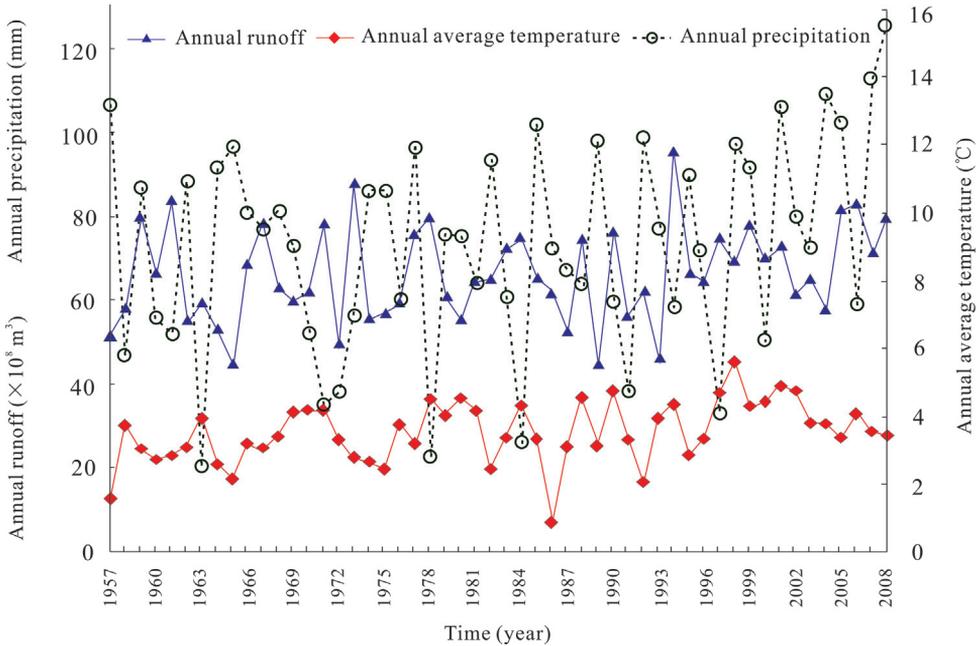

**Fig. 1.** Hydrological and climatic observation data of Yarkland river basin, Northwest China.

The five scales of time are designated as $s_1$ to $s_5$, the Fig. 2(a) presents five variation patterns of AR. The $s_1$ curve retains a large amount of residual from the raw data, and drastic fluctuations exist in the study period. These characteristics indicate that, although the runoff varied greatly throughout the study period, there was a hidden increasing trend. The $s_2$ curve still retains a considerable amount of residual, as indicated by the presence of 4 peaks and 4 valleys. However, the $s_2$ curve is much smoother than the $s_1$ curve, which allows the hidden increasing trend to be more apparent. The $s_3$ curve retains much less residual, as indicated by the presence of 2 peaks and 2 valleys. Compared to $s_2$, the increase in runoff over time is more apparent in $s_3$. Finally, the $s_5$ curve presents an ascending tendency, whereas the increasing trend is obvious in the $s_4$ curve. Fig. 2(b) and (c) present five variation patterns of AAT and AP, which show the similar variation patterns to AR at the five scales of time.

Based on the results of wavelet analysis, five linear regression equations to describe the hydro-climate relationships between AR with AAT and AP were fitted at the five scales of time (Table 1), which show multi-time scale responses of annual runoff to regional climatic change in the Yarkand River basin of Northwest China.

Table 1 tells us that the significant level of wavelet regression equations (WREs) at the five scales of time from $s_1$ (2-year scale) to $s_5$ (32-year scale) achieved as high as $\alpha = 0.001$, and the WRE at 1-year scale are also achieved the significant level of $\alpha = 0.01$. By comparing their AIC values, we also compare the fitting effects of WREs at different time scales. The order of fitting effect of WREs at different time scales is as follows: the fitting effect of the WRE at $s_5$ (32-year scale) is the best, that at $s_4$ (16-year scale) is second, that at $s_3$ (8-year scale) is third, that at $s_4$ (4-year scale) is fourth, that at $s_1$ (4-year scale) is the penult, and that at $s_0$ is the worst.

## Summary

Combining wavelet analysis and regression method, we developed an integrated approach, the wavelet regression (WR), which can be used to show the multi-time scale responses of a hydrological

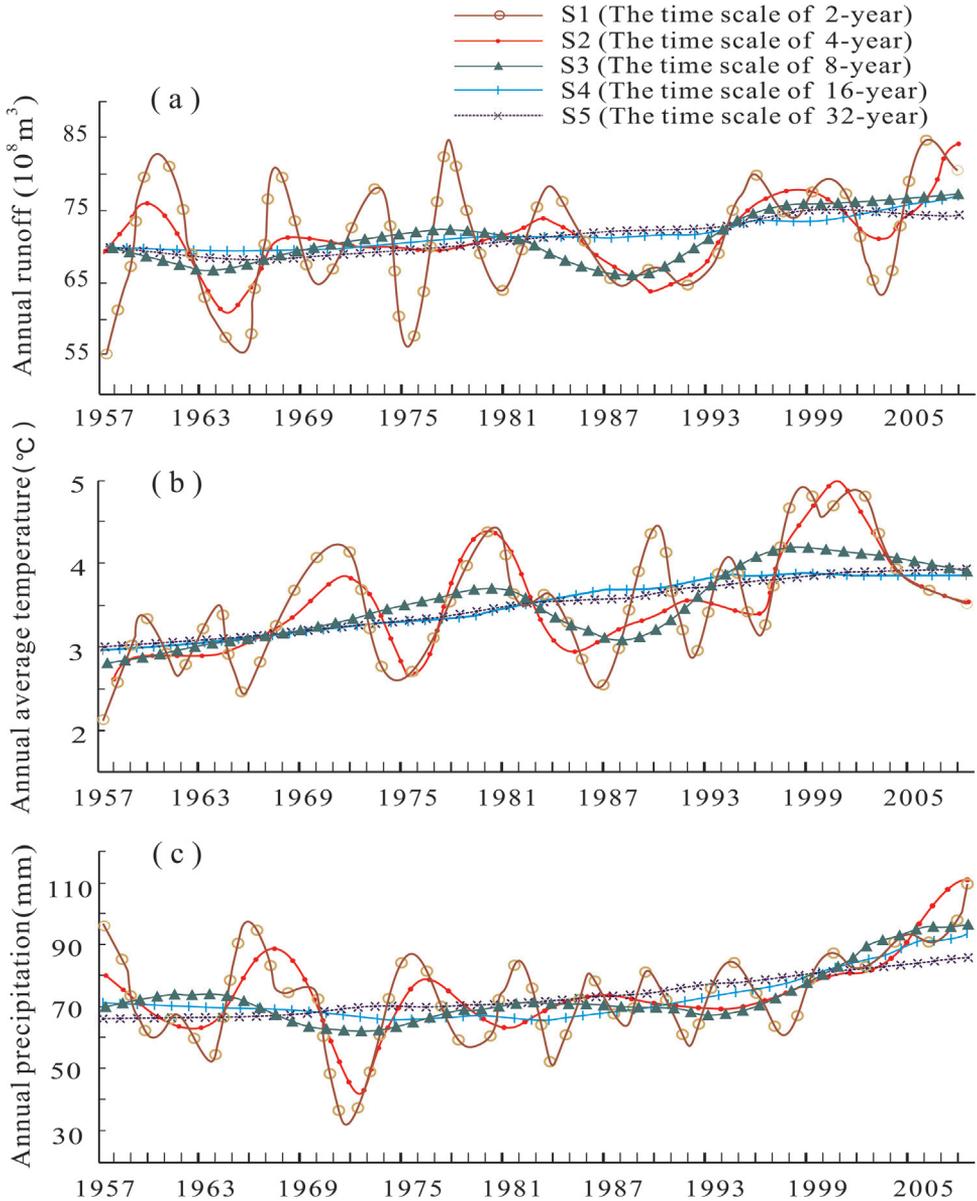

**Fig. 2.** Variation patterns at different time scales of (a) annual runoff, (b) annual average temperature, and (c) annual precipitation.

variable to climate change. The principle of the approach is that the wavelet analysis is used to reveal the variation patterns of a hydrological variable and its related climatic factors at different time scales, and then the regression method is used to show the hydro-climate relationship between the hydrological variable with its related climatic factors for each pattern at the chosen time scale. To illustrate the application of the approach, the hydro-climate relationships between annual runoff (AR)

**Table 1**
Wavelet regression equations to describe the relationships between annual runoff and annual average temperature & annual precipitation at different time scales.

| Time scale | Regression equation | $R^2$ | F | Significance level $\alpha$ | AIC |
|---|---|---|---|---|---|
| $s_0$ | $AR = 8.8150AAT + 0.0793AP - 9.6203$ | 0.3666 | 12.7340 | 0.01 | 400.528 |
| $s_1$ | $AR = 10.7510AAT + 0.0742AP - 26.5651$ | 0.5765 | 29.9478 | 0.001 | 209.924 |
| $s_2$ | $AR = 11.9308AAT + 0.0706AP - 36.9412$ | 0.6754 | 45.7753 | 0.001 | 143.263 |
| $s_3$ | $AR = 11.4623AAT + 0.1397AP - 39.2254$ | 0.7991 | 87.5066 | 0.001 | 12.960 |
| $s_4$ | $AR = -10.6715AAT + 0.6335AP + 113.1979$ | 0.9509 | 425.6954 | 0.001 | -96.714 |
| $s_5$ | $AR = 25.6166AAT - 0.2411AP - 130.8705$ | 0.9999 | 1598.985 | 0.001 | -209.172 |

*Notes: AR* – annual runoff, *AAT* – annual average temperature, and *AP* – annual precipitation; $s_0$, $s_1$, $s_2$, $s_3$, $s_4$ and s5 represent 1-year, 2-year, 4-year, 8-year, 16-year and 32-year scale, respectively.

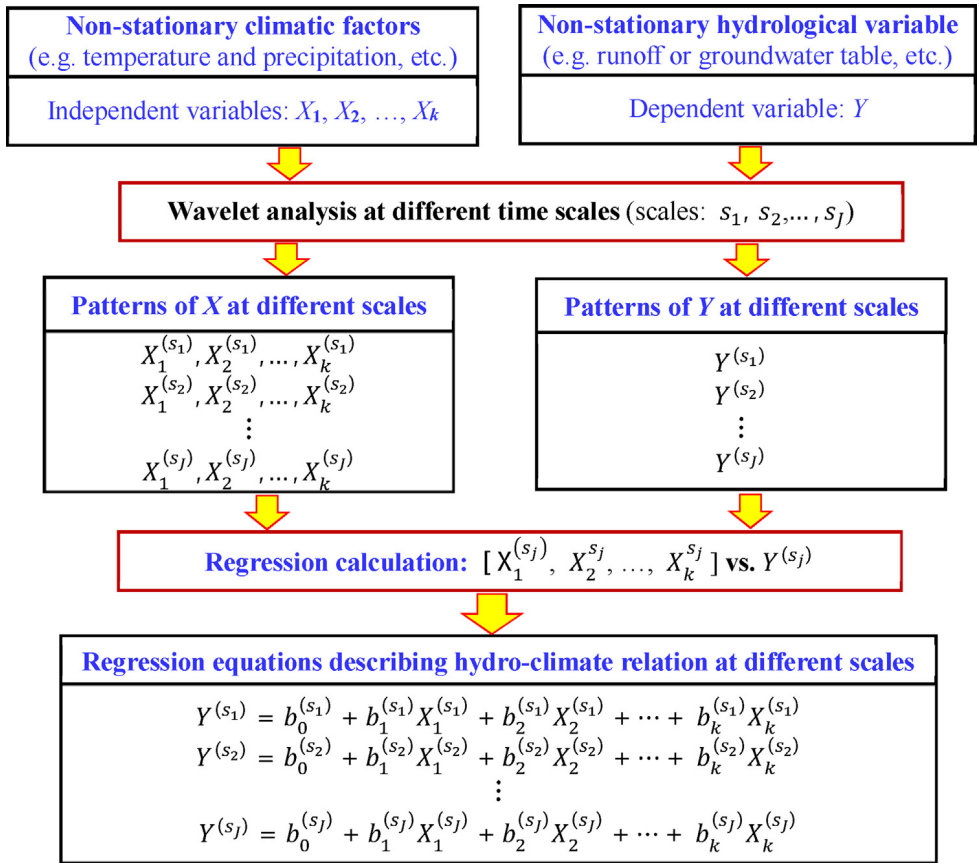

**Fig. 3.** The procedure of wavelet regression for multi-time scale analyses of hydro-climate relationship.

with annual average temperature (AAT) and annual precipitation (AP) in the Yarkand River basin of Northwest China have been analyzed at different time scales. Though the effects of wavelet regression equations (WREs) at time scales are different (the larger time scale, the better effect), overall the wavelet regression equations (WREs) can significantly describe the relations between AR with AAT and AP based at chosen time scales.

## Additional information

Previous studies showed that hydro-climate processes are complex systems [18–22], and the observed data of climate and hydrology are non-stationary and stochastic [23–26]. How can we find out the patterns hidden in the stochastic and non-stationary data? Multi-time scale analysis is an approach to attempt [27,28].

To show the multi-time scale responses of a hydrological variable (e.g. runoff, evaporation, or groundwater level, etc.) to climate change, we developed an integrated approach by combining wavelet analysis and regression method, which is called wavelet regression (WR). The main idea of the approach is that the wavelet analysis is used to reveal the variation patterns of a hydrological variable and its related climatic factors at different time scales, and then the regression method is used to show the hydro-climate relationship between the hydrological variable and its related climatic factors for each pattern at the chosen time scale. The procedure of this approach is shown as Fig. 3, and the effectiveness of the approach has been verified in some case studies [14–17,29–32].

## Conflicts of interest

The author declares that they have no conflicts of interest.

## Acknowledgements

This work was supported by the National Natural Science Foundation of China [41630859]; the Strategic Priority Research Program of the Chinese Academy of Sciences [Grant No. XDA19030204]; and the Open Foundation of State Key Laboratory, Desert and Oasis Ecology, Xinjiang Institute of Ecology and Geography, Chinese Academy of Sciences [No. G2014-02-07].